%% file: main.tex
\documentclass[a4paper,onecolumn,draftcls]{IEEEtran}
\newcommand\beq{\begin{equation}}
\newcommand\eeq{\end{equation}}
	
	\usepackage{graphicx}
	\usepackage{amsmath,amssymb}
	\usepackage{amsfonts}
	\usepackage{color}
	\usepackage{cite}
    \usepackage{bm}

\usepackage{cite}
\usepackage{graphicx}

\usepackage{algorithm}
\usepackage{algpseudocode}
\usepackage{steinmetz}

\newcommand{\eqdef}{\triangleq}
\newcommand{\trasp}{\text{T}}
\newcommand{\herm}{\text{H}}

\newcommand{\gammab}{\pmb{\gamma}}
\newcommand{\x}{\mathbf{x}}

\renewcommand{\a}{\mathbf{a}}

\newcommand{\proj}{\mathcal{P}}

\newcommand{\Sigmab}{\boldsymbol{\Sigma}}

\newcommand{\Rset}{\mathbb{R}}
\newcommand{\Cset}{\mathbb{C}}

\input{arxiv}

\begin{document}


\title{Conformal Reconfigurable Intelligent Surfaces: A Cylindrical Geometry Perspective}

\author{Filippo~Pepe,
Ivan~Iudice,
Giuseppe~Castaldi,
Marco~Di~Renzo,
Vincenzo~Galdi
\thanks{This work was partially supported by the European Union-Next Generation EU under the Italian National Recovery and Resilience Plan (NRRP), Mission 4, Component 2, Investment 1.3, CUP E63C22002040007, partnership on ``Telecommunications of the Future'' (PE00000001 - program ``RESTART''). The work of M. Di Renzo was supported in part by the European Union through the Horizon Europe project COVER under grant agreement number 101086228, the Horizon Europe project UNITE under grant agreement number 101129618, the Horizon Europe project INSTINCT under grant agreement number 101139161, and the Horizon Europe project TWIN6G under grant agreement number 101182794, as well as by the Agence Nationale de la Recherche (ANR) through the France 2030 project ANR-PEPR Networks of the Future under grant agreement NF-PERSEUS 22-PEFT-004, and by the CHIST-ERA project PASSIONATE under grant agreements CHIST-ERA-22-WAI-04 and ANR-23-CHR4-0003-01.}}

\maketitle
\acceptednotice

\begin{abstract}
Curved reconfigurable intelligent surfaces (RISs) represent a promising frontier for next-generation wireless communication, enabling adaptive wavefront control on nonplanar platforms such as unmanned aerial vehicles and urban infrastructure.  
This work presents a systematic investigation of cylindrical RISs, progressing from idealized surface-impedance synthesis to practical implementations based on simple one-bit meta-atoms. Exact analytical and geometrical-optics-based models are first developed to explore fundamental design limits, followed by a semi-analytical formulation tailored to discrete, reconfigurable architectures. 
This model enables efficient beam synthesis using both evolutionary optimization and low-complexity strategies, including the minimum power distortionless response method, and is validated through full-wave simulations. Results confirm that one-bit RISs can achieve directive scattering with manageable sidelobe levels and minimal hardware complexity. These findings establish the viability of cylindrical RISs and open the door to their integration into dual-use wireless platforms for real-world communication scenarios.
\end{abstract}


\section{Introduction}

Reconfigurable intelligent surfaces (RISs) and metasurfaces have emerged as promising technologies for future wireless communication systems, offering unprecedented control over electromagnetic (EM) wave propagation across the microwave, millimeter-wave, and  terahertz domains \cite{Bas2019,Cheng:2022ri,Malevich:2025vl}. By dynamically adjusting the reflection properties of subwavelength elements, these engineered surfaces can tailor the wireless channel in real time, enabling functions such as coverage enhancement, interference suppression, and improved link reliability in environments where infrastructure is sparse, obstructed, or highly dynamic \cite{Liu:201ri}. 

Owing to their low-power, lightweight, and programmable nature, RISs are particularly suited to dual-use scenarios, encompassing both civilian and tactical applications. Platforms such as unmanned aerial vehicles (UAVs) \cite{Li2021} and connected vehicles \cite{Naaz:2024et} can exploit RIS-assisted communication to improve spectral efficiency and maintain robust connectivity under non-line-of-sight conditions. In dense urban environments, RISs can enhance vehicle-to-everything, UAV-to-infrastructure, and UAV-to-vehicle links, where multipath and shadowing effects often degrade performance \cite{Kisselef:2020ri}. More broadly, the smart radio environment paradigm \cite{DiR2020} envisions RISs as passive yet reconfigurable elements that reshape the wireless channel itself, enabling advanced functions such as multi-hop relaying and wavefront control without relying on traditional active relays. These advantages make RISs attractive for flexible deployment across both mobile platforms and fixed infrastructure, including in disaster recovery, temporary coverage extension, and secure battlefield communications.

Most existing RIS implementations rely on {\em planar} configurations, which are often poorly suited for real-world structures. Platforms such as UAV fuselages, vehicle bodies, and infrastructure elements like lampposts, traffic poles, and support towers typically involve {\em curved} surfaces. Mounting flat RIS panels on these platforms can introduce mechanical difficulties and degrade EM performance. This motivates growing interest in {\em conformal} RIS platforms that can be integrated directly onto the geometry of the host structure.

In the EM research community, conformal metasurfaces have been extensively explored as versatile platforms for manipulating waves on curved geometries, with applications ranging from optics to microwaves. On the modeling side, advances such as the conformal boundary optics framework have extended generalized sheet transition conditions to free-form surfaces, enabling accurate treatment of curvature-induced phase accumulation and modified boundary conditions \cite{Wu:2018mo}. At optical frequencies, flexible all-dielectric metasurfaces have enabled lensing, cloaking, and wavefront shaping on curved carriers, with experiments showing good tolerance to mechanical deformation and variations in incidence angle \cite{Cheng:2016ad}. Recent reviews on flexible and conformal metasurfaces \cite{Zhou:2024fm} have emphasized practical fabrication aspects, including stretchable substrates, transfer printing, and integration on complex three-dimensional (3D) platforms, while also noting challenges related to alignment, robustness under bending, and long-term reliability.

In the microwave and millimeter-wave domains, conformal metasurfaces have been widely investigated for beam steering and radiation control. Early work demonstrated active cylindrical metasurface lenses that use varactor tuning for wide-angle steering \cite{Li:2019wa}. Later studies showed that passive conformal reflective metasurfaces can enhance beam manipulation and gain in 5G millimeter-wave multi-input multi-output configurations \cite{Malik:2024bs}. Transmissive conformal metasurface antennas implemented on arbitrary cylindrical supports further confirmed the feasibility of practical curved apertures \cite{Liu:2023ct}. At terahertz frequencies, beam steering on flexible curved substrates has been shown to maintain the desired deflection performance across a broad range of bending radii \cite{Shiri:2023tb}. Additional progress includes data-driven, real-time microwave reconfigurability based on adaptive reflective surfaces \cite{Wen:2023rd} and reconfigurable spin-decoupled conformal metasurfaces fabricated on 3D-printed curved substrates \cite{Fu:2025rs}. Recent developments also encompass conformal holographic metasurfaces for scattering-pattern modulation \cite{Zhang:2024am}, multifunctional time-modulated conformal designs for radar and spectrum-control applications \cite{Sun:2024af,Sun:2024fc}, reconfigurable free-form camouflage metasurfaces \cite{Li:2018rf}, and conformal reconfigurable holographic metasurfaces for multifunctional radiation and scattering manipulation \cite{Zhang:2025cr}. Collectively, these results demonstrate significant opportunities offered by conformal platforms while also highlighting practical challenges associated with curvature-dependent performance, fabrication tolerances, and packaging on nonplanar carriers.

From a communication-oriented perspective, the influence of RIS geometry has begun to receive systematic attention \cite{Mizmizi:2023cm}. Comparative analyses of planar, linear, and cylindrical RIS topologies \cite{Cui:2023io} indicate that curvature affects illumination uniformity, effective aperture, and near-field interactions. These effects translate into measurable differences in link budget, signal-to-noise ratio, and outage probability. Such observations underscore the importance of modeling approaches that connect EM behavior, hardware constraints, and communication-level performance metrics.

Within this broader context, conformal RIS platforms emerge as a particularly attractive solution. By following the natural curvature of the supporting structure, conformal RISs enable seamless integration on real-world platforms while introducing additional geometrical degrees of control, which can be exploited to enhance beam steering, improve coverage shaping, and increase spatial diversity \cite{An:2025fi,Mursia:2025t3}. In this sense, curvature itself becomes a functional design parameter, enabling more flexible control of radiation patterns and wavefront shaping than is attainable with conventional planar RIS implementations.

\begin{figure}
    \centering
    \includegraphics[width=.8\linewidth]{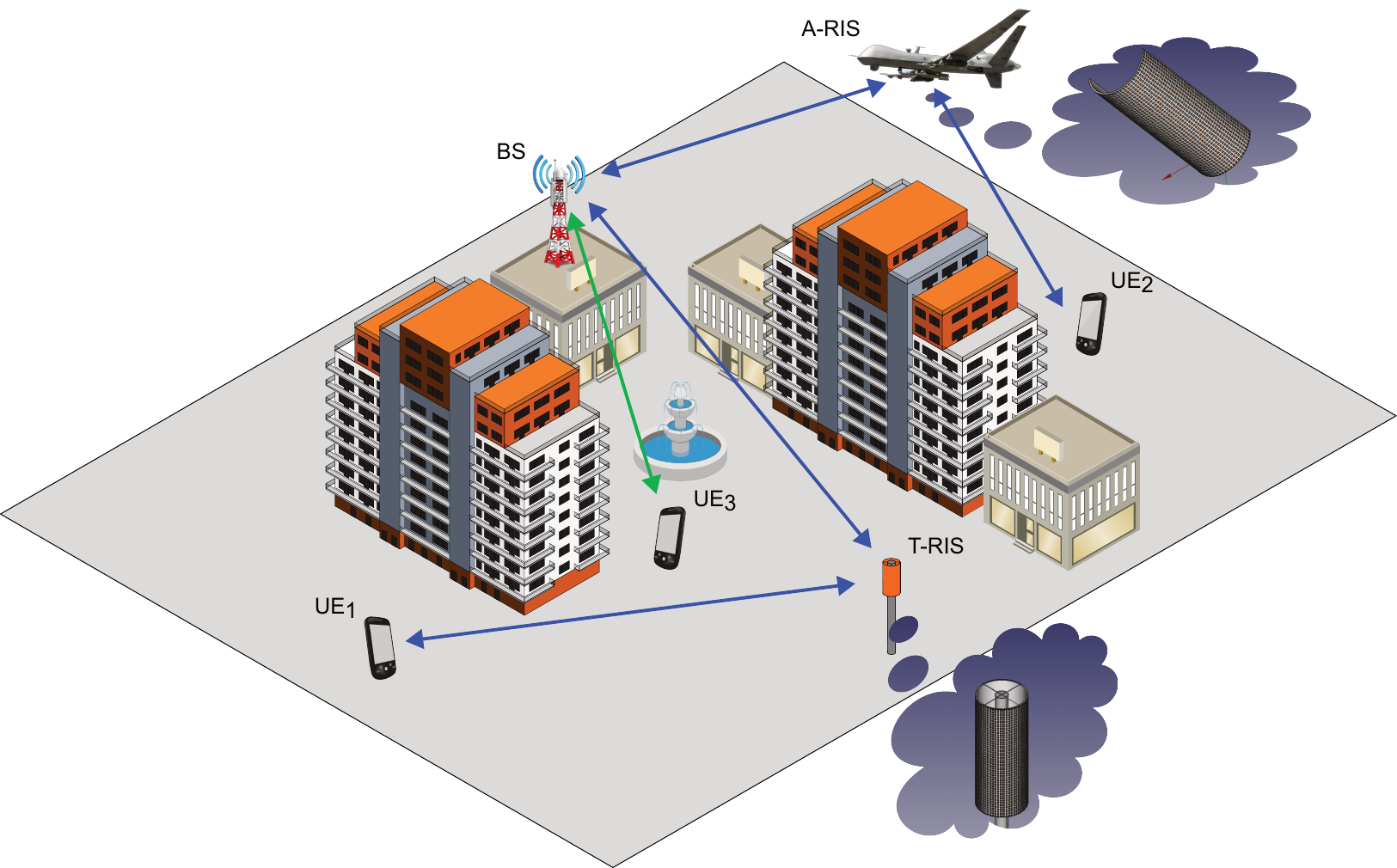}
    \caption{Illustrative scenario of conformal RIS deployment in a mixed aerial (A-RIS) and terrestrial (T-RIS) communication network. Cylindrical RISs are integrated onto curved surfaces, such as UAV fuselages and lampposts, to enhance coverage and enable dynamic beam control in complex propagation environments. A base station (BS) communicates with multiple user equipments (UE$_1$, UE$_2$, UE$_3$), with RISs assisting in overcoming line-of-sight obstructions and improving link quality.}
    \label{fig:Figure1}
\end{figure}

In this work, we focus on the {\em cylindrical} geometry, which provides a simple yet representative case for conformal RIS integration. Its regular shape facilitates both modeling and fabrication, and it is relevant to many real-world platforms. As illustrated in Figure \ref{fig:Figure1}, cylindrical RISs can be deployed on UAV fuselages and on urban infrastructure such as poles and lampposts. These implementations support functions such as adaptive beamforming and passive relaying while preserving the structural and visual characteristics of the host platform.

We investigate an idealized 2D cylindrical RIS, encompassing both continuous surface-impedance models and practical digitally coded, column-controlled architectures. By integrating analytical, semi-analytical, and full-wave EM modeling, we assess beam-steering performance while systematically examining the effects of curvature and phase quantization. The synthesis problem is addressed through a combination of optimization strategies, including evolutionary algorithms and low-complexity approaches based on the minimum power distortionless response (MPDR) criterion.
Compared with previous studies on conformal or cylindrical RISs, this work introduces a unified framework that connects exact cylindrical impedance synthesis, a locally passive formulation, semi-analytical modeling of discrete one-bit meta-atoms with angular dependence, optimization-based beam synthesis, and full-wave validation. This multi-level approach provides a clearer picture of the fundamental and practical limits governing beam steering on cylindrical RIS platforms, while highlighting tradeoffs between hardware simplicity and achievable EM performance.

%

\section{Results and Discussion}

%
\begin{figure}
	 \centering
		\includegraphics[width=.6\linewidth]{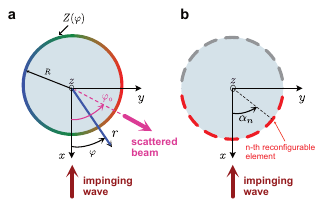}
		\caption{Problem geometry. a) Idealized scenario with surface impedance. The 2D Cartesian and associated cylindrical reference systems are also shown.
			b) Realistic scenario with space-sampled and phase-quantized reconfigurable elements. The elements directly illuminated by the incident wave are shown in red, while those located in the shadow region are shown in gray.}
			\label{fig:Figure2}
\end{figure}

\subsection{Problem Geometry and Formulation}
Referring to Figure \ref{fig:Figure2}, we consider a 2D scenario involving an infinitely long cylinder aligned along the $z$-axis, with radius $R$, embedded in free space. The structure is illuminated by a plane wave with time-harmonic dependence $\exp(j \omega t)$, incident normally to the cylinder axis. Due to the translational invariance along $z$, all fields are independent of this coordinate. Without loss of generality, we assume the wave propagates along the negative $x$-direction, with a $z$-polarized electric field of unit amplitude, expressed as
\begin{equation}
	E_{iz}(r, \varphi) = \exp(j k_0 x) = \exp(j k_0 r \cos\varphi),
	\label{eq:Eiz}
\end{equation}
where $k_0 = \omega / c_0 = 2\pi / \lambda_0$ is the wavenumber, and $c_0$ and $\lambda_0$ denote the speed of light and wavelength, respectively.

The objective is to design the EM response of the cylindrical surface in order to redirect the scattered field into a narrow beam centered around a desired angle $\varphi_o$.

\subsection{Idealized Scenario: Surface Impedance}
To explore the fundamental physical limitations of cylindrical metasurfaces, in analogy with the planar case studied in Ref. \cite{DiRenzo:2022cm}, we begin by analyzing an idealized, isotropic scenario. Specifically, we consider a cylinder of radius $R$, characterized by a continuous, scalar surface impedance distribution $Z(\varphi)$, as illustrated in Figure \ref{fig:Figure2}a. This model imposes the following boundary condition at $r=R$ \cite{Hoppe:1995ib}:
\begin{equation}
	E_z(R,\varphi) = Z(\varphi) H_{\varphi}(R,\varphi),
	\label{eq:IBC}    
\end{equation}
where $E_z$ and $H_{\varphi}$ denote the total $z$-polarized electric field and $\varphi$-polarized magnetic field, respectively.

This problem can be addressed analytically using a Bessel-Fourier expansion. The scattered electric field can be expressed as \cite{Harrington:2001th}
\begin{equation}
	E_{sz}\left(r,\varphi\right)=\sum_{m=-\infty}^{\infty}c_m 
	\exp[-j m\left(\varphi-\pi/2\right)]
	\mathcal{H}_m^{(2)}\!\!\left(k_0r\right),
	\label{eq:Esz1}
\end{equation}
where $\mathcal{H}_m^{(2)}(\cdot)$ is the $m$th-order Hankel function of the second kind  \cite{Abramowitz:1965ho}, and $c_m$ are  expansion coefficients.

To design the scattered field to resemble a locally planar wavefront directed toward a desired angle $\varphi_o$, we enforce the following condition at the surface $r = R$:
\begin{equation}
	E_{sz}(R, \varphi) \sim \exp\left[-j k_0 R \cos(\varphi - \varphi_o)\right].
\label{eq:PW}
\end{equation}

Using the Jacobi-Anger expansion \cite{Abramowitz:1965ho}, the corresponding coefficients in Equation (\ref{eq:Esz1}) can be determined as:
\begin{equation}
	c_m = \frac{\exp[j m (\varphi_o - \pi)] \mathcal{J}_m(k_0R)}{\mathcal{H}_m^{(2)}(k_0R)},
\end{equation}
where $\mathcal{J}_m(\cdot)$ is the $m$th-order Bessel function of the first kind \cite{Abramowitz:1965ho}.

The associated magnetic fields, for both incident and scattered components, can be obtained from Maxwell’s curl equations:
\begin{equation}
	H_{\nu \varphi}(r, \varphi) = \frac{1}{j \omega \mu_0} \frac{\partial E_{\nu z}(r, \varphi)}{\partial r},\quad \nu=i,s,
\end{equation}
with $\mu_0$ denoting the free-space permeability. Substituting into the boundary condition (\ref{eq:IBC}), the required surface impedance is given by:
\begin{eqnarray}
	\frac{Z(\varphi)}{\eta_0} &=& \frac{E_{iz}(R, \varphi) + E_{sz}(R, \varphi)}{\eta_0 \left[H_{i\varphi}(R, \varphi) + H_{s\varphi}(R, \varphi)\right]}\nonumber\\
	&=&\frac{
		1 + \exp(-j k_0 R \cos \varphi) \displaystyle{\sum_{m=-\infty}^{\infty}} c_m \exp[-j m\left(\varphi-\pi/2\right)] \mathcal{H}_m^{(2)}(k_0R) 
	}{
		\cos\varphi  - j \exp(-j k_0 R \cos \varphi) \displaystyle{\sum_{m=-\infty}^{\infty}} c_m 
		\exp[-j m\left(\varphi-\pi/2\right)]
		\dot{\mathcal{H}}_m^{(2)}(k_0R) 
	},
		\label{eq:ZZ}
\end{eqnarray}
where $\eta_0$ is the free-space intrinsic impedance, and the overdot indicates differentiation with respect to the argument.
As is well known, the infinite series in Equation (\ref{eq:ZZ}) can be truncated to a finite number of terms $M \sim k_0 R$, representing the number of available EM degrees of freedom \cite{Bucci:1989ot}. 

In principle, the coefficients $c_m$ can be tailored via Fourier synthesis to synthesize arbitrary far-field patterns, beyond simple beam steering. The synthesis procedure can be summarized in two main steps:
\begin{enumerate}
	\item Expand the desired far-field pattern into a Fourier series to obtain the coefficients $c_m$;
	\item Compute the corresponding surface impedance distribution using Equation \eqref{eq:ZZ}.
\end{enumerate}

However, as previously observed in the planar case \cite{DiRenzo:2022cm}, this approach generally ensures only {\em global} passivity, without guaranteeing {\em local} passivity. The resulting impedance profile in Equation (\ref{eq:ZZ}) is, in general, {\em complex-valued}, indicating that different regions of the surface may exhibit gain or loss, even if the overall structure may remain passive on average.

\begin{figure}
	\centering
	\includegraphics[width=.8\linewidth]{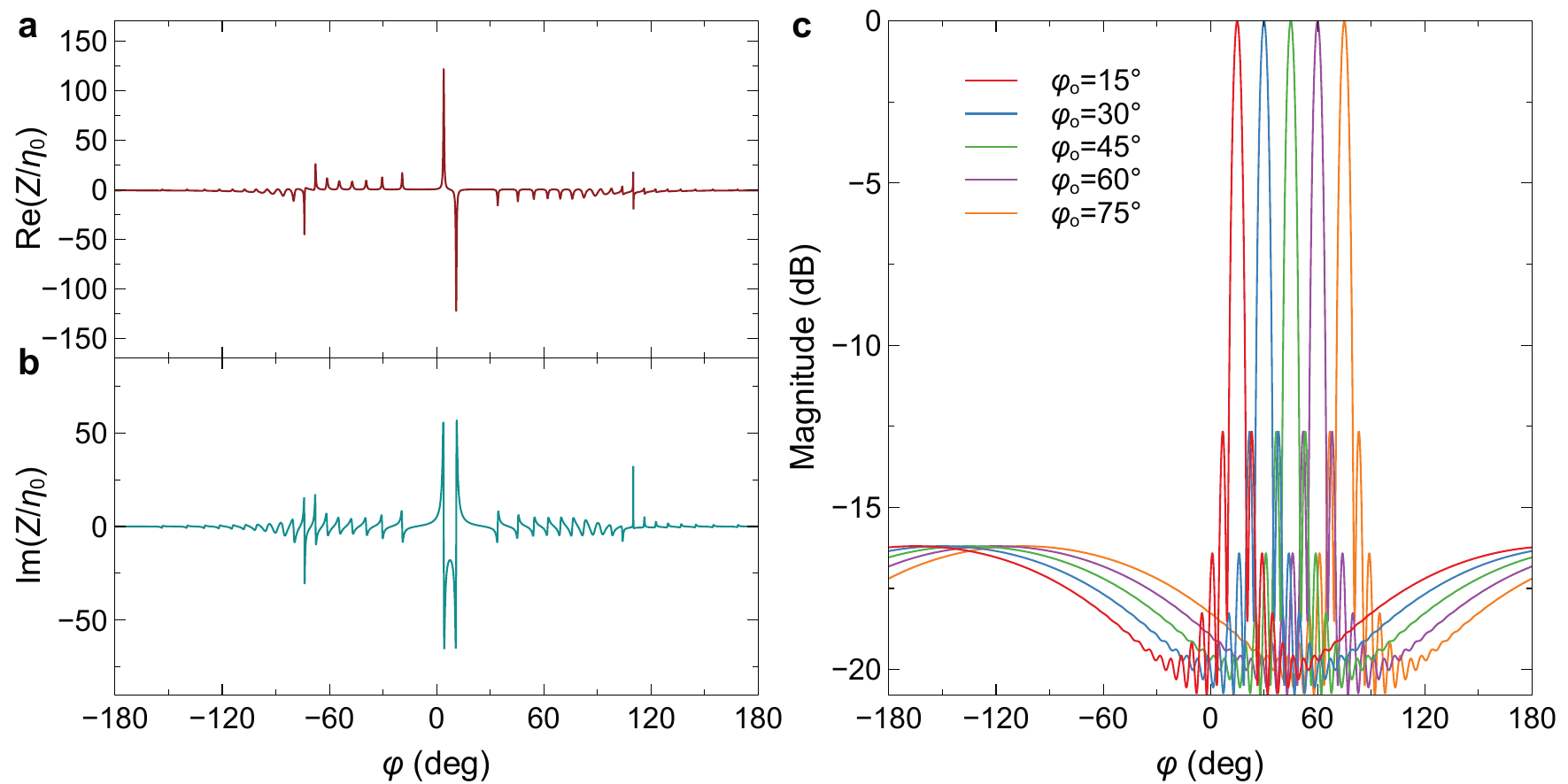}
	\caption{
		Exact analytical synthesis. a,b) Real and imaginary parts, respectively, of the normalized surface impedance computed from Equation (\ref{eq:ZZ}) for beam steering toward $\varphi_o=15^\circ$.
		 c) Far-field scattering patterns for different values of the steering angle $\varphi_o$. Results are normalized with respect to the maximum value.}
	\label{fig:Figure3}
\end{figure}

Figure \ref{fig:Figure3} presents representative numerical results. In particular, Figure \ref{fig:Figure3}a,b show the normalized surface impedance computed from Equation (\ref{eq:ZZ}) for a steering angle of $\varphi_o = 15^\circ$. As expected, the surface resistance exhibits both positive and negative values, indicating the coexistence of locally lossy and active (gain) regions. Similar behaviors are observed for other steering angles (not shown for brevity). Figure \ref{fig:Figure3}c displays the far-field scattering patterns for steering angles ranging from $15^\circ$ to $75^\circ$ in $15^\circ$ increments. The patterns remain consistent in shape and peak amplitude across the range of steering angles, while maintaining effective sidelobe level (SLL) control. Notably, forward scattering is significantly suppressed, a behavior reminiscent of parity-time-symmetric invisibility cloaking mechanisms \cite{Hao:2019ua}, where a tailored balance of gain and loss is used to minimize the total scattering signature.

%

To ensure local passivity of the structure, an alternative approach can be adopted. This method approximates the cylindrical surface locally by its tangent plane and applies the well-established geometrical-optics (GO) framework developed for planar anomalous reflectors \cite{Sun:2012he}. A detailed derivation is provided in the Appendix. This formulation yields a purely imaginary surface impedance given by
\begin{equation}
	Z(\varphi)=-\frac{j\eta_0}{\cos\varphi}\cot\left[\frac{\Phi_r(\varphi)}2\right],
	\label{eq:Zlocal}
\end{equation}
where
\begin{equation}
	\Phi_r\left(\varphi\right) =   k_0R \left[\cos(\varphi-\varphi_o)+\cos\varphi\right].
	\label{eq:Phir}
\end{equation}

This approach inherits the known limitations of GO-based methods in the synthesis of  planar anomalous reflectors \cite{Diaz:2017ft}. Moreover, due to the tangent-plane approximation, its validity is further restricted to configurations with electrically large curvature radii, where $k_0 R \gg 1$.

\begin{figure}
	\centering
	\includegraphics[width=.8\linewidth]{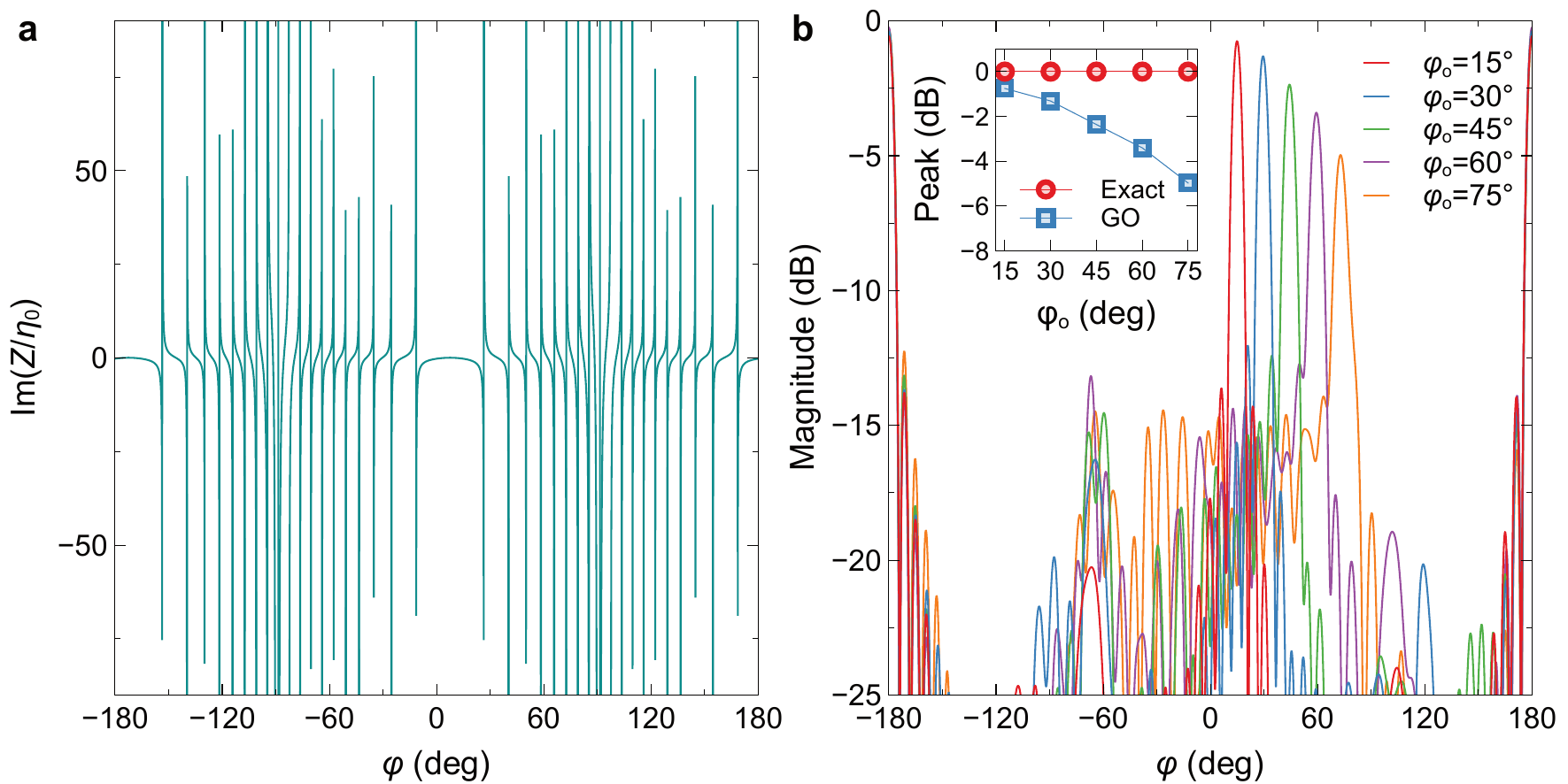}
	\caption{
		GO synthesis. a) Purely imaginary normalized surface impedance computed from Equation (\ref{eq:Zlocal}) for beam steering toward $\varphi_o=15^\circ$.
		b) Far-field scattering patterns for different values of the steering angle $\varphi_o$. Results are normalized with respect to the overall maximum value. The inset compares the peak values with those obtained via the exact synthesis (Figure \ref{fig:Figure3}) as a function of the steering angle $\varphi_o$.}
	\label{fig:Figure4}
\end{figure}

Figure~\ref{fig:Figure4} presents representative results for the locally passive synthesis strategy. Figure \ref{fig:Figure4}a shows the purely imaginary normalized surface impedance obtained from Equation~\eqref{eq:Zlocal} for a steering angle of $\varphi_o = 15^\circ$. The reactance profile exhibits rapid spatial fluctuations and a wide dynamic range, highlighting the complexity involved in the surface design required to achieve accurate beam shaping. Figure \ref{fig:Figure4}b displays the corresponding far-field scattering patterns for different steering angles. Compared to the exact synthesis results in Figure \ref{fig:Figure3}, two key differences are observed.
First, in the absence of active and lossy regions, forward scattering is no longer suppressed, leading to pronounced lobes near $\varphi = 180^\circ$. Second, the beam-steering efficiency deteriorates as the steering angle increases. This effect is quantified in the inset of Figure \ref{fig:Figure4}b, which compares the main-beam amplitude achieved by the two approaches as a function of the steering angle, revealing a degradation of up to nearly 5 dB.

This trend is consistent with observations in the planar case \cite{Diaz:2017ft}. In that context, it was shown that globally passive designs, based on a balance of gain and loss, can be effectively synthesized using fully passive structures by carefully engineering strong spatial dispersion. While this approach could, in principle, be extended to cylindrical geometries, its implementation would be considerably more complex. Furthermore, spatially dispersive structures are inherently difficult to reconfigure, which limits their practicality in dynamic RIS applications.

%

\subsection{Realistic Scenario: Reconfigurable Digital Elements}

Both  idealized approaches discussed earlier require surface-impedance profiles with rapid spatial variations and large dynamic ranges, which pose significant challenges for implementation on low-complexity, reconfigurable platforms. To address this, we consider a more practical scenario, illustrated in Figure \ref{fig:Figure2}b, in which the cylindrical metasurface is composed of identical elements (``meta-atoms'') with reconfigurable reflection properties \cite{Cui:2014cm}. 

For the generic $n$-th element, we assume the reflection coefficient $\gamma_n$ can take values from a discrete set of digital states, 
$\gamma_n \in \Gamma_n \eqdef \left\{\gamma_n^{(\ell)}\right\}_{\ell=1}^L$,
where $L = 2^b$ denotes the number of available states and $b$ is the number of control bits. Importantly, since the elements are illuminated from different angles along the cylindrical surface, the set of accessible states generally differs between elements, i.e., $\Gamma_{n_1} \neq \Gamma_{n_2}$ for $n_1 \neq n_2$. This angular dependence is incorporated into the model by computing the directional response of each element through full-wave simulations under the assumption of a planar, periodic array configuration.

Note that the assumed 2D scenario implies that the element response varies only along the $\varphi$-direction and remains constant along the cylinder axis ($z$-direction). This simplification is a reasonable approximation for many practical column-controlled RIS configurations, where a single biasing line is used to control all elements in a column, thereby reducing electronic complexity.

Unlike the idealized cases discussed earlier, this scenario does not admit a closed-form analytical solution. A rigorous treatment would require full-wave numerical methods, such as finite-element or finite-difference techniques. However, these approaches are computationally intensive and impractical for optimization tasks that involve repeated evaluations of the scattering response. To address this, we develop a semi-analytical, approximate model that offers significantly reduced computational cost while retaining sufficient accuracy for design and optimization. The validity of this model is later confirmed through comparison with full-wave simulations.

Our model is based on a physical-optics approximation \cite{Osipov:2017me} and is inspired by the theory of conformal (circular) arrays \cite{Persson:2006th}. Within this framework, the far-field response is approximated as
\begin{equation}
	F(\varphi) \propto \lim_{r\rightarrow\infty}  \sqrt{r} E_{sz}(r,\varphi)
	=\sum_{n=1}^N \gamma_n a_n(\varphi),
	\label{eq:FFphi}
\end{equation}
where
\begin{equation}
	a_n(\varphi) = E_{n}(\varphi) \exp
	\left\{
	jk_0R\left[\cos(\varphi - \alpha_n)+\cos\alpha_n\right]
	\right\}
	\label{eq:an}
\end{equation}
represents the contribution of the $n$-th element.
Here, $\alpha_n$ denotes the angular position of the element on the cylindrical surface (see Figure \ref{fig:Figure2}b), and $E_n(\varphi)$ describes its scattering pattern, modeled as a cosine function relative to the local surface normal. The summation in Equation (\ref{eq:FFphi}) is limited to those elements directly illuminated by the incident wavefront (shown in red color in Figure \ref{fig:Figure2}b).

In principle, the model described above can be extended to more general geometries with variable curvature, as long as the fundamental assumptions of the physical-optics approximation remain valid. In particular, multiple reflections, which are not accounted for in the current formulation, should be negligible for the approximation to hold \cite{Osipov:2017me}.

To enable a synthesis strategy that jointly controls beam steering, beamwidth, and SLL, we define the angular exclusion set
$S_\varphi \eqdef \{\varphi \in [-\pi,\pi] : |\varphi-\varphi_0| > \Delta \varphi/2\}$
where $\Delta \varphi$ denotes the desired beamwidth.

We reformulate the far-field expression in Equation \eqref{eq:FFphi} using matrix-vector notation as
\begin{equation}
	F(\varphi) = \a^\trasp(\varphi) \, \gammab,
	\label{eq:far-field}
\end{equation}
where 
$\a(\varphi) \eqdef [a_1(\varphi), a_2(\varphi),
\ldots, a_N(\varphi)]^\trasp \in \Cset^N$
is the steering vector, and 
$\gammab \eqdef [\gamma_1, \gamma_2,
\ldots, \gamma_N]^\trasp \in \Gamma$
collects the reflection coefficients. The feasible domain $\Gamma$ is defined as the Cartesian product $\Gamma_1 \times \Gamma_2 \times \ldots \times \Gamma_N$, with each $\Gamma_n$ representing the discrete set of available states for the $n$-th element.

The far-field power pattern is then given by
\begin{equation}
	|F(\varphi)|^2 = |\a^\trasp(\varphi) \gammab|^2 = \gammab^\herm \a^*(\varphi) \a^\trasp(\varphi) \gammab
	\label{eq:far-field-power},
\end{equation}
where $(\cdot)^\trasp$, $(\cdot)^\herm$, and $(\cdot)^*$ denote the transpose, Hermitian, and complex conjugate operators, respectively.

A widely adopted strategy in array synthesis focuses on minimizing the SLL. This design objective can be formulated as the following discrete optimization problem:
\begin{equation}
\begin{split}
	\min_{\{\gamma_n\}_{n=1}^N} \quad &
	\frac{\|F(\varphi)\|_{\mathcal{L^\infty(S_\varphi)}}}{\|F(\varphi)\|_{\mathcal{L^\infty}}} \\
	\text{s.t.} \quad & \gamma_n \in \Gamma_n, \forall n \in \{1,2,\ldots,N\},
\end{split}
\label{eq:sll-prob}
\end{equation}
where $\|F(\varphi)\|{\mathcal{L}^\infty(S\varphi)} \triangleq \max_{\varphi \in S_\varphi} |F(\varphi)|$ denotes the maximum SLL outside the main beam, and $\|F(\varphi)\|{\mathcal{L}^\infty} \triangleq \max_{\varphi \in [-\pi, \pi[} |F(\varphi)|$ corresponds to the global peak.

The formulation in Equation (\ref{eq:sll-prob}) constitutes a mixed-integer fractional quadratic programming problem, which is generally challenging to solve efficiently due to 
its non-convex and combinatorial nature \cite{Lee2011}. 
When the solution space becomes large, i.e., when $|\Gamma| = L^N$ grows, an exhaustive search (ES) quickly becomes computationally infeasible. As a result, alternative approaches such as relaxation techniques and evolutionary algorithms are commonly employed to find near-optimal solutions \cite{Ata2020,Li2021,Lin2020,Yue2023,Che2024}.

In our study, we evaluated both genetic algorithms (GAs) \cite{Haupt:2007ga} and low-complexity methods, such as the MPDR beamforming technique \cite{Cap1969}. The latter, detailed in the Appendix, yields a closed-form solution. As an additional low-complexity alternative, we also considered a brute-force spatial sampling and phase quantization of the GO solution given in Equation (\ref{eq:Zlocal}), which likewise results in a closed-form expression.

Note that the choice of beamwidth $\Delta\varphi$ is not entirely arbitrary, as it is physically constrained by the electrical size of the structure, i.e., the number of EM degrees of freedom available \cite{Bucci:1989ot}. Intuitively, smaller structures produce less directional scattering, while larger ones can support narrower beams. As a result, assigning an excessively small beamwidth may lead to an ill-conditioned synthesis problem. As a physically meaningful reference, we adopt the beamwidth of a uniform circular array with the same size and number of elements \cite{Persson:2006th}.

In what follows, we consider a representative configuration consisting of a cylindrical surface with radius $R=40$ cm, resulting in $N=30$ illuminated elements. The system operates at a carrier frequency of $3.6$ GHz with one-bit resolution ($b=1$). These parameters are consistent with typical UAV-assisted communication scenarios defined within the 5G New Radio standard. 

As a reconfigurable meta-atom, we adopt the design proposed in Ref. \cite{Yang:2016ap}, consisting of a square metallic patch connected to the ground plane through a vertical via controlled by a positive-intrinsic-negative (PIN) diode. The geometry is scaled appropriately for operation at the target frequency. Figure \ref{fig:FigureS1} shows the meta-atom structure and its angular response, obtained through full-wave simulations at 3.6 GHz (see the Methods section for mode details). This response is used in the semi-analytical model of Equation (\ref{eq:FFphi}) to define the local reflection coefficients.

As can be observed in Figure S1, the reflection magnitude remains stable and close to unity across the range of incidence angles, while the phase difference between the two states is exactly $180^\circ$ at normal incidence but gradually deteriorates at oblique angles. Consequently, meta-atoms located near the edges of the illuminated region become less effective in switching between states, thereby reducing the number of available degrees of control. Although this particular design is not optimized for angular stability, and improved configurations have been proposed in the literature \cite{Shabanpour:2024eo}, the phase difference will inherently vanish as the incidence approaches grazing.

%
\begin{figure}
	\centering
	\includegraphics[width=\linewidth]{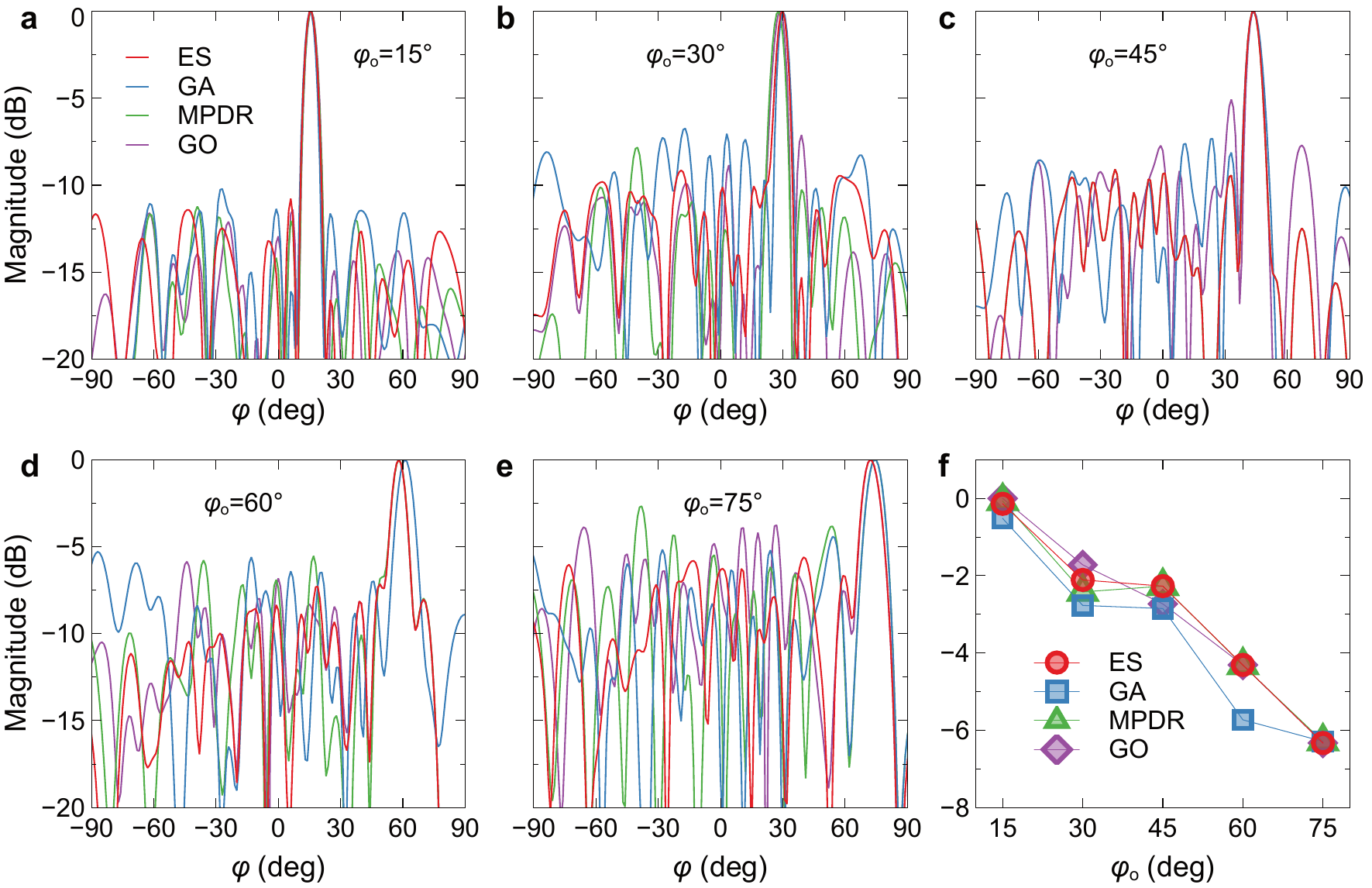}
	\caption{a,b,c,d,e) Examples of synthesized scattering patterns for a cylindrical RIS operating at 3.6 GHz, with radius $R=40$ cm, $N=30$ illuminated elements, and one-bit ($b=1$) resolution. The patterns correspond to steering angles $\varphi_o=15^\circ$, $30^\circ$, $45^\circ$, $60^\circ$, and $75^\circ$, respectively, and are normalized to their individual peak values.
		f) Comparison of the main beam levels at the nominal steering angle across the various synthesis methods, as the steering direction is varied. Levels are normalized with respect to the maximum value observed at $\varphi_o=15^\circ$.}
	\label{fig:Figure5}
\end{figure}

Figure \ref{fig:Figure5} summarizes representative results from the considered synthesis approaches. In particular, Figures \ref{fig:Figure5}a–e show the synthesized scattering patterns for steering angles ranging from $15^\circ$ to $75^\circ$ in  $15^\circ$ increments, with a beamwidth intentionally set as $20\%$ larger than the reference value, obtained using ES, GA, MPDR, and GO. For ease of comparison, especially in terms of SLL, each pattern is normalized to its peak value. To evaluate efficiency trends as a function of steering angle, Figure \ref{fig:Figure5}f plots the main-beam levels at the nominal direction, normalized to the maximum obtained at $\varphi_o = 15^\circ$, for all synthesis methods.

Several observations emerge. All methods perform similarly in the main-lobe region, with only minor steering offsets. The most significant differences appear in the SLL, where the GO method typically performs worst. This is expected given its underlying approximations and the use of coarse sampling and quantization. The MPDR method often yields satisfactory results, occasionally outperforming the GA, and in some cases matching the ES solution (for example, $\varphi_o = 45^\circ$ in Figure \ref{fig:Figure5}c). As expected, performance generally degrades with increasing steering angle. Figure \ref{fig:Figure5}f shows that this degradation follows a trend similar to the idealized case (see inset in Figure \ref{fig:Figure4}b), regardless of the synthesis method. This suggests that certain limitations stem from the physical configuration itself, and in particular from the constraint of local passivity.

It is worth noting that the ES method does not always yield the highest peak at the nominal direction (e.g., $\varphi_o = 30^\circ$ in Figure \ref{fig:Figure5}f). This is because the optimization problem in Equation (\ref{eq:sll-prob}) targets sidelobe suppression rather than directly maximizing on-axis gain, which may result in slight mispointing.

In terms of computational effort, the MPDR and GO methods are highly efficient, as they rely on closed-form expressions and require negligible processing time. By contrast, the GA approach, implemented as described in the Methods section, required approximately ten minutes of runtime on a dedicated workstation. The ES method  was significantly more demanding, taking nearly five days to complete on the Turing supercomputing facility at the Italian Aerospace Research Centre.

Overall, the MPDR method offers a good balance between performance and computational cost, making it a promising candidate even for higher-bit configurations.

%
\begin{figure}
	\centering
	\includegraphics[width=.8\linewidth]{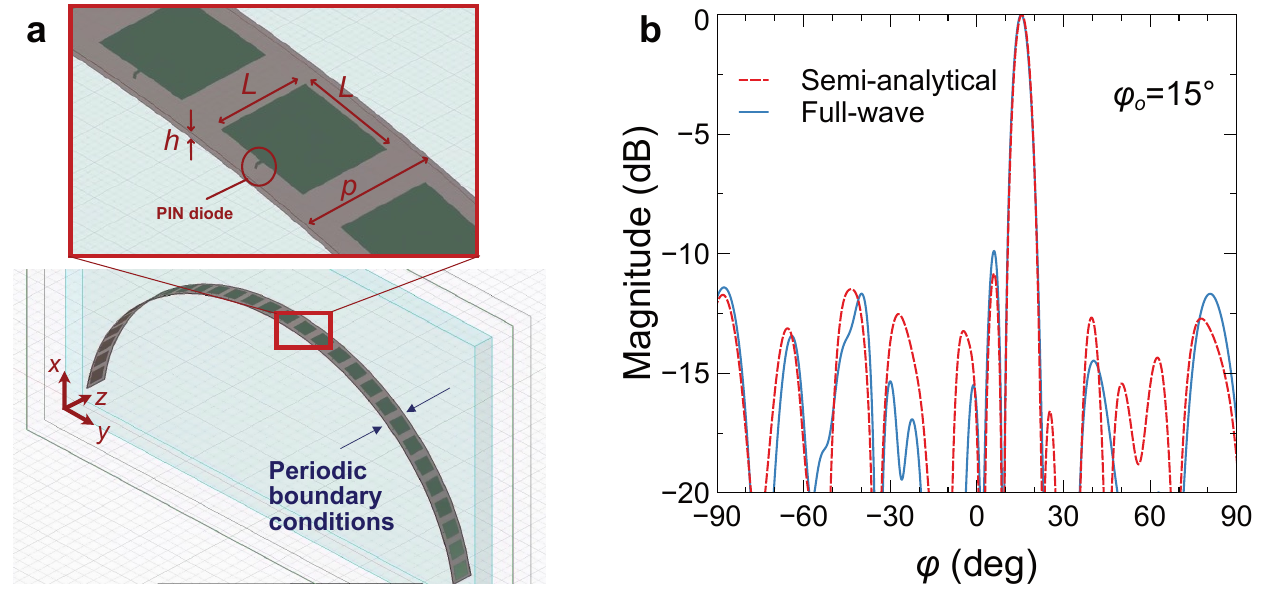}
	\caption{a) Geometry of the simulated cylindrical RIS, with a magnified inset highlighting  structural details (for a complete description of the meta-atom, see the inset of Figure \ref{fig:FigureS1}).
		b) Comparison of scattering patterns obtained using the semi-analytical model and full-wave simulations, for an ES synthesis targeting a steering angle of $\varphi_o = 15^\circ$. Patterns are normalized to their individual peak values.}
	\label{fig:Figure6}
\end{figure}

\subsection{Full-Wave Validation}

To close the loop, we validate the semi-analytical model by comparing its results with finite-element full-wave simulations (see the Methods section for details).

Representative results are shown in Figure \ref{fig:Figure6}. Specifically, Figure \ref{fig:Figure6}a illustrates the simulated geometry, while Figure \ref{fig:Figure6}b compares the scattering patterns obtained from the semi-analytical model and full-wave simulations, for an ES-based synthesis with a steering angle of $\varphi_o = 15^\circ$. The results show good overall agreement, confirming the validity of the semi-analytical approach in capturing the main characteristics of the radiation pattern. The main beam directions and shapes are well matched, with only minor discrepancies in the SLL. 

Qualitatively similar trends are observed for other steering angles, as shown in Figure \ref{fig:FigureS2}. The discrepancies in SLL remain within about 3 dB for moderate steering angles up to $60^\circ$, and increase to approximately 5 dB for the more extreme case of 75°. These deviations stem from the simplifying assumptions inherent to the semi-analytical model, including locally planar approximations and the neglect of higher-order EM interactions, which are fully captured in the full-wave simulations. It is also worth noting that, in the full-structure numerical model, the meta-atoms are not treated as planar replicas but are geometrically conformed to the cylindrical surface. Therefore, the intrinsic distortion associated with the 2D-to-3D mapping of the meta-atoms, along with the resulting curvature-induced modifications of the local current distribution and effective electrical size, is inherently accounted for. The reasonable agreement observed between the two approaches thus indicates that the impact of conformal distortion on the scattering performance is already properly reflected in the numerical validation.

Each full-wave simulation required on the order of two hours per scattering pattern on a dedicated workstation, which renders this approach impractical for direct use within an optimization loop involving repeated pattern evaluations. By contrast, the semi-analytical model provides results in a negligible fraction of that time, typically within seconds. Despite the small to moderate discrepancies observed with respect to full-wave data, this substantial reduction in computational cost enables rapid exploration of the design space and supports efficient direct synthesis techniques such as MPDR.

\subsection{Additional Considerations}
The proposed cylindrical RIS is designed for operation at a single frequency where the two meta-atom states exhibit an ideal 180$^\circ$ degree phase difference under normal incidence. 
To assess the impact of frequency detuning, we carried out additional simulations, with representative results reported in Figure \ref{fig:FigureS3}. As the operating frequency departs from its design value of 3.6 GHz, the reflection magnitude of the meta-atom in both ON and OFF states under normal incidence (Figure S3a) exhibits only mild variations, whereas the phase contrast between the two states (Figure S3b) degrades more noticeably. As shown in Figure S3c, this mainly affects the sidelobe structure of the scattered field, while the main beam direction and amplitude remain comparatively stable over the considered frequency range. This behavior is consistent with the narrowband nature of the selected meta-atom, which was not designed for dispersion or angular stability. Wider operational bandwidths could be achieved by employing meta-atoms specifically engineered for frequency-stable phase response, such as multi-resonant or dispersion-flattened designs reported in the literature \cite{Simovski:2025bo}. Integrating such elements into the present cylindrical RIS architecture is feasible and represents a natural direction for future developments.

Regarding the influence of element density, the dimensions used in this work are consistent with typical RIS implementations at 3.6 GHz and reflect practical constraints imposed by UAV-scale cylindrical platforms. Although smaller reconfigurable meta-atoms on the order of $\lambda_0/5$ have been demonstrated \cite{Dai:2025jc} and could, in principle, increase the number of available degrees of control, this benefit must be balanced against the added hardware complexity and the reduced area available for biasing networks. It is also important to note that beam-pointing accuracy is mainly limited by phase quantization, whereas cell spacing mainly affects sidelobe behavior once near-Nyquist sampling is satisfied. Further reduction in meta-atom size would make diode integration and bias routing more challenging. Within these considerations, the chosen element dimensions provide an effective compromise between practical implementability and beamforming performance for cylindrical RIS deployments.

Finally, in the present study, the polarization sensitivity is essentially dictated by the adopted meta-atom geometry rather than by the cylindrical conformal arrangement itself. In particular, the asymmetric loading associated with the PIN diode breaks the unit-cell symmetry, resulting in a polarization-selective response. The cylindrical geometry mainly affects the local angle of incidence, while the polarization response remains governed by the meta-atom design. It is worth noting that polarization-insensitive metasurfaces have been widely reported (see, e.g., \cite{Li:2024mp}). These approaches could be directly extended to cylindrical RISs to mitigate polarization sensitivity. A systematic investigation of polarization effects in conformal RISs, including cross-polar coupling induced by curvature, will be addressed in future work.

\section{Conclusion}

This study presented an investigation of cylindrical RISs, combining analytical, semi-analytical, and full-wave methods to evaluate their beam-steering capabilities. Starting from ideal surface-impedance models, we derived exact synthesis conditions and explored both globally and locally passive designs. 

A key outcome of this work is the demonstration that directive scattering can be achieved with extremely simple hardware. We focused on one-bit reconfigurable meta-atoms based on binary switching and showed that, when combined with suitable synthesis strategies, they can deliver robust beam control. Among the approaches evaluated, the MPDR method proved particularly attractive due to its closed-form expression and low computational cost. 
The semi-analytical model underlying the synthesis was validated against full-wave simulations, confirming its accuracy in capturing the main beam features and showing only minor discrepancies in SLL for moderate steering angles.

Future work will explore several directions, including higher-bit reconfigurable designs and the incorporation of space-time coding strategies, originally developed for planar RISs \cite{Zhang:2018st}, into conformal architectures for enhanced spatial and spectral control. On the experimental side, we are progressing toward prototype fabrication and validation, with laboratory measurements to be followed by field tests on UAV platforms within a dedicated flight arena under construction at the Italian Aerospace Research Centre.

From a methodological standpoint, the analytical synthesis developed in this work leverages the separability of cylindrical coordinates. Although such separability does not hold for more general curved geometries, the semi-analytical and numerical frameworks introduced here do not rely on this constraint and can be extended to non-cylindrical conformal RISs. Advancing these generalizations, which will require a full vector formulation on more general coordinate systems, is a current focus of our ongoing research.

Overall, the results highlight that low-complexity RIS architectures can enable effective beam steering even in non-planar configurations. This opens the door to practical integration on curved surfaces such as UAV fuselages and infrastructure poles, supporting a wide range of dual-use wireless communication scenarios.


\section{Methods}
\subsection{Numerical Simulations} 
Full-wave simulations are performed using the finite-element-based commercial software Ansys HFSS \cite{HFSS}. Two types of simulations are carried out: meta-atom characterization and full-structure analysis.

\subsubsection{Meta-atom characterization} 
The meta-atom geometry used for characterization is shown in the inset of Figure \ref{fig:FigureS1}. The square metallic patch has side length $L = 26$ mm and is printed on a grounded substrate of thickness $h = 1.57$ mm, with a lattice period of $p = 38$ mm. The substrate is modeled using its nominal material parameters, namely a relative permittivity $\varepsilon_r = 2.2$ and a loss tangent $\tan\delta = 0.001$. The patch is connected to the ground plane through a metallic via incorporating a PIN diode, which enables binary reconfiguration. The diode is modeled as a lumped-element series circuit, consisting of an RL pair (1.5 $\Omega$, 0.7 nH) in the ON state and an LC pair (0.7 nH, 0.15 pF) in the OFF state \cite{Yang:2016ap}.
All metals are treated as perfect electric conductors. To emulate an infinitely periodic array, the unit cell is truncated along the transverse directions using phase-shift (Floquet) boundary conditions. 

A wave port located at a distance of $1.5\lambda_0$ above the unit cell is used to excite the structure with a plane wave having a $z$-polarized electric field, over a range of incidence angles. 
The corresponding scattering parameters are extracted for the two states of the PIN diode. 
 
An adaptive tetrahedral meshing strategy is employed. Convergence is controlled by enforcing a Maximum Delta Energy criterion of 0.05, so that the adaptive process terminates when the estimated variation in stored energy between consecutive mesh refinements falls below this threshold. A minimum of ten adaptive passes is imposed to ensure robust convergence. The initial mesh is defined with a minimum element size of 0.2 mm in the vicinity of the PIN diode region and 2 mm over the metallic patch. Upon convergence, the final mesh comprises approximately 55,000 elements.

\subsubsection{Cylindrical RIS simulations}
For the full-structure analysis, we adopt the geometry shown in Figure \ref{fig:Figure6}a, which consists of an array of identical meta-atoms conformally wrapped on a cylindrical surface. Unlike the planar configuration used in the unit-cell simulations, each meta-atom is geometrically curved so that its patch, via, and substrate follow the local cylindrical profile. The full 3D model is generated automatically through a script written in the HFSS scripting language to ensure consistent geometry and parameter control. 

The cylindrical RIS has a radius of $R=40$ cm and is embedded in an air-filled parallelepiped of dimensions $56.66 \times 88.33 \times 3.8$ cm$^3$. To emulate an infinitely long structure, periodic boundary conditions are applied along the cylinder axis ($z$-direction). Along the transverse directions ($x$ and $y$), a perfectly matched layer with a thickness of 8.33 cm is used to simulate an open region.

The structure is illuminated by a normally incident plane wave with the same $z$-polarized electric field assumed in the analytical formulation. Far-field quantities are extracted through HFSS post-processing, with angular scattering patterns exported using a one-degree sampling resolution unless otherwise specified.

An adaptive tetrahedral meshing strategy is employed to ensure numerical accuracy. Convergence is controlled by imposing a Maximum Delta Energy of 0.1, so that the adaptive process terminates once the estimated energy variation between consecutive mesh passes falls below this threshold. The initial mesh is defined with a minimum element length of 0.4 mm in the vicinity of the PIN diode and 5 mm on the metallic patch, while all other initial mesh settings are kept at their default values. The final converged mesh consists of approximately 1.7 million elements.

\subsection{Optimization}
The ES method minimizes the objective function in Equation (\ref{eq:sll-prob}) by exhaustively searching all possible combinations of PIN diode states across the RIS. The GA-based optimization is carried out using an in-house code implemented in MATLAB \cite{MATLAB}, with the goal of minimizing the maximum field magnitude outside the desired angular window. The GA is configured with a population size of 1000, 200 generations, a crossover probability of 0.9, and a mutation probability of 0.05. The MPDR method, which provides a closed-form low-complexity solution, is described in detail in the Appendix.

\appendix

\section{Details on  GO Synthesis}
Assuming an incident plane-wave electric field as in Equation (\ref{eq:Eiz}), and approximating the cylindrical surface locally by its tangent plane, we prescribe a reflected field at position $(R, \varphi)$ that is directed toward a desired angle $\varphi_o$, given by Equation (\ref{eq:PW}).

Following the approach used for planar anomalous reflectors \cite{Sun:2012he}, the local reflection coefficient can be approximated as
\begin{equation}
	\Gamma(\varphi)\equiv\frac{E_{rz}(R,\varphi)}{E_{iz}(R,\varphi)}\simeq\exp\left\{-jk_0R\left[\cos(\varphi-\varphi_o)+\cos\varphi\right]\right\}.
	\label{eq:R1}
\end{equation}

Alternatively, the reflection coefficient can be expressed in terms of the surface impedance as \cite{Sun:2012he}
\begin{equation}
	\Gamma(\varphi)\simeq\frac{Z(\varphi)-Z_w(\varphi)}{Z(\varphi)+Z_w(\varphi)},
    \label{eq:R2}
\end{equation}
where the wave impedance of the incident field is \cite{Harrington:2001th}
\begin{equation}
	Z_w(\varphi)=\frac{\eta_0}{\cos\varphi}.
\end{equation}

By equating the expressions in Equations (\ref{eq:R1}) and (\ref{eq:R2}) and solving for $Z(\varphi)$, we obtain the surface-impedance formula given in Equation (\ref{eq:Zlocal}), with the phase function defined in Equation (\ref{eq:Phir}).

\section{Details on  MPDR Synthesis}
To reduce the computational complexity of the synthesis process, we propose a formulation based on the MPDR beamforming strategy  \cite{Cap1969}. The problem is cast as a discrete optimization task:
\begin{equation}
	\begin{split}
		\min_{\{\gamma_n\}_{n=1}^N} \quad &
		\|F(\varphi)\|^2_{\mathcal{L}^2} \\
		\text{s.t.} \quad & F(\varphi_o) = \rho \, \exp\left(j\psi\right) \\
		& \gamma_n \in \Gamma_n, \forall n \in \{1,2,\ldots,N\}.
	\end{split}
	\label{eq:moe-prob-con}
\end{equation}

Here, $\|F(\varphi)\|^2_{\mathcal{L}^2} \eqdef \int_{-\pi}^{\pi} |F(\varphi)|^2 \, d\varphi$ denotes the total scattered power, and the constraint enforces a prescribed field amplitude $\rho$ and phase $\psi$ in the desired steering direction $\varphi_o$. The discrete nature of the design is accounted for by restricting each $\gamma_n$ to a finite set of quantized reflection states $\Gamma_n$.

Note that Equation (\ref{eq:moe-prob-con}) employs the quadratic norm, in contrast to Equation (\ref{eq:sll-prob}), which uses the uniform norm. From a physical perspective, the quadratic norm evaluated over the interval $[-\pi, \pi[$ corresponds to the total power scattered by the RIS, providing a meaningful measure of the overall scattering efficiency.

Using Equations (\ref{eq:far-field}) and (\ref{eq:far-field-power}), the optimization problem in Equation (\ref{eq:moe-prob-con}) can be reformulated as
\begin{subequations}
\label{eq:moe-prob-mat}
\begin{alignat}{3}
	\min_{\gammab} \quad &
	\gammab^\herm \Sigmab \, \gammab \\
	\text{s.t.} \quad &  \a^\trasp(\varphi_o) \, \gammab = \rho \, \exp\left(j\psi\right)
	\label{eq:moe-lin-cons} \\
	& \gammab \in \Gamma,
	\label{eq:moe-discrete-cons}
\end{alignat}
\end{subequations}
where $\Sigmab \eqdef \int_{-\pi}^\pi \a^*(\varphi)
\a^\trasp(\varphi) \, d\varphi \in \Cset^{N \times N}$
is a Hermitian matrix 
which can be evaluated either analytically or numerically, depending on the specific RIS configuration.

By relaxing the constraint in Equation \eqref{eq:moe-discrete-cons} and allowing $\gammab \in \Cset^N$, the problem in \eqref{eq:moe-prob-mat} becomes a linearly constrained quadratic programming problem. This type of problem admits a {\em closed-form} solution \cite{Cap1969}, given by
\begin{equation}
	\gammab_{\text{MPDR}} = \lambda \, \Sigmab^{-1} \a^*(\varphi_o) \, \exp\left(j\psi\right),
	\label{eq:MPDR}
\end{equation}
where $\lambda \in \Rset$ is a scalar chosen to satisfy the linear constraint in Equation \eqref{eq:moe-lin-cons}, and is given by
\begin{equation}
	\lambda = \frac{\rho}{\a^\trasp(\varphi_o) \Sigmab^{-1} \a^*(\varphi_o)}.
	\label{eq:lambda}
\end{equation}

To recover a feasible solution to the original discrete optimization problem in Equation \eqref{eq:moe-prob-mat}, we introduce a projection operator based on a minimum-distance quantization strategy. Specifically,
\begin{equation}
	\proj(\gammab) = \arg\min_{\x \in \Gamma}
	\|\x - \gammab\|^2_2,
	\label{eq:proj-complex}
\end{equation}
where $\gammab \in \Cset^N$.

When the discrete reflection coefficients satisfy $\left|\gamma_n^{(\ell)}\right| \approx 1$ for all $n \in \{1,2,\ldots,N\}$ and $\ell \in \{1,2,\ldots,L\}$, the projection outcome becomes largely insensitive to the amplitude of the continuous vector $\gammab$. As a result, the scalar constant $\lambda$ can be omitted without loss of generality.

However, the output of the projection is significantly influenced by the phase term $\psi$. To account for this, the optimal RIS configuration can be determined by solving the following 1D optimization problem:
\begin{equation}
	\min_{\psi \in [-\pi, \pi[}
	\|\a^\trasp(\varphi) \, \overline{\gammab}(\psi)\|^2_{\mathcal{L}^2(S_\varphi)},
\end{equation}
where
\begin{equation}
	\begin{split}
		\|\a^\trasp(\varphi) \, \overline{\gammab}(\psi)\|^2_{\mathcal{L}^2(S_\varphi)}
		& \eqdef \int_{S_\varphi} |\a^\trasp(\varphi) \, \overline{\gammab}(\psi)|^2 \, d\varphi \\
		& = \overline{\gammab}^\herm(\psi) \, \Sigmab_{S_\varphi} \, \overline{\gammab}(\psi).
	\end{split}
\end{equation}
Here, $\overline{\gammab}(\psi) \eqdef
\proj\left[\Sigmab^{-1} \a^*(\varphi_o) \, e^{j\psi}\right]$,
and $\Sigmab_{S_\varphi} \in \Cset^{N \times N}$
is given by $\int_{S_\varphi} \a^*(\varphi)
\a^\trasp(\varphi) \, d\varphi$.

\section{Additional Results}

\begin{figure}
	\centering
	\includegraphics[width=.5\linewidth]{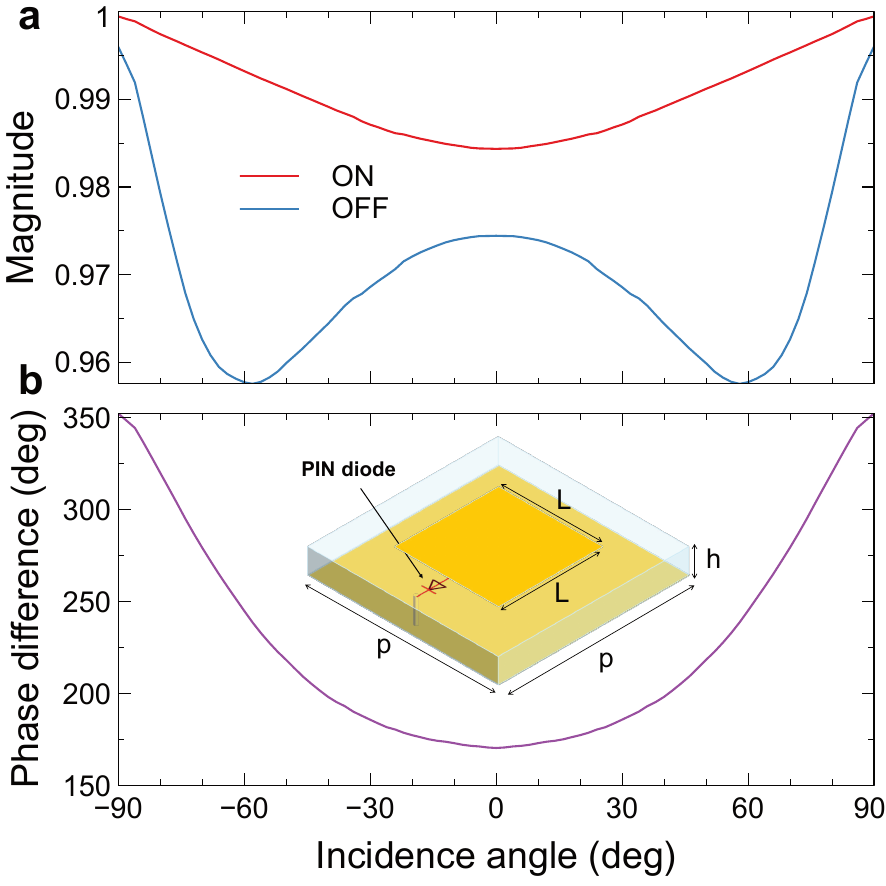}
	\caption{a, b) Magnitude and phase difference of the reflection coefficient for the one-bit reconfigurable meta-atom, evaluated at the operating frequency of 3.6 GHz as a function of the incidence angle, for both operational states. The geometry of the meta-atom is illustrated in the inset of panel (b). Design parameters are: $L=26$ mm, $p=38$mm, and $h=1.57$mm. The substrate has a relative permittivity $\varepsilon_r=2.2$ and a loss tangent $\tan\delta=0.001$.}
	\label{fig:FigureS1}
\end{figure}

Figure \ref{fig:FigureS1} presents the magnitude and phase responses of the one-bit reconfigurable meta-atom at the operating frequency of 3.6 GHz, plotted as a function of the incidence angle for the two states of the PIN diode. The meta-atom geometry is shown in the inset of panel (b), with all relevant geometrical and material parameters provided in the figure caption.

As discussed in the main text, the reflection magnitude remains nearly constant and close to unity over the full range of incidence angles. However, the phase difference between the two states, which is exactly $180^\circ$ at normal incidence, gradually degrades as the angle becomes more oblique. This effect reduces the switching effectiveness of meta-atoms positioned near the edges of the illuminated region, leading to a corresponding decrease in the number of effective degrees of freedom.
 
\begin{figure}
	\centering
	\includegraphics[width=.8\linewidth]{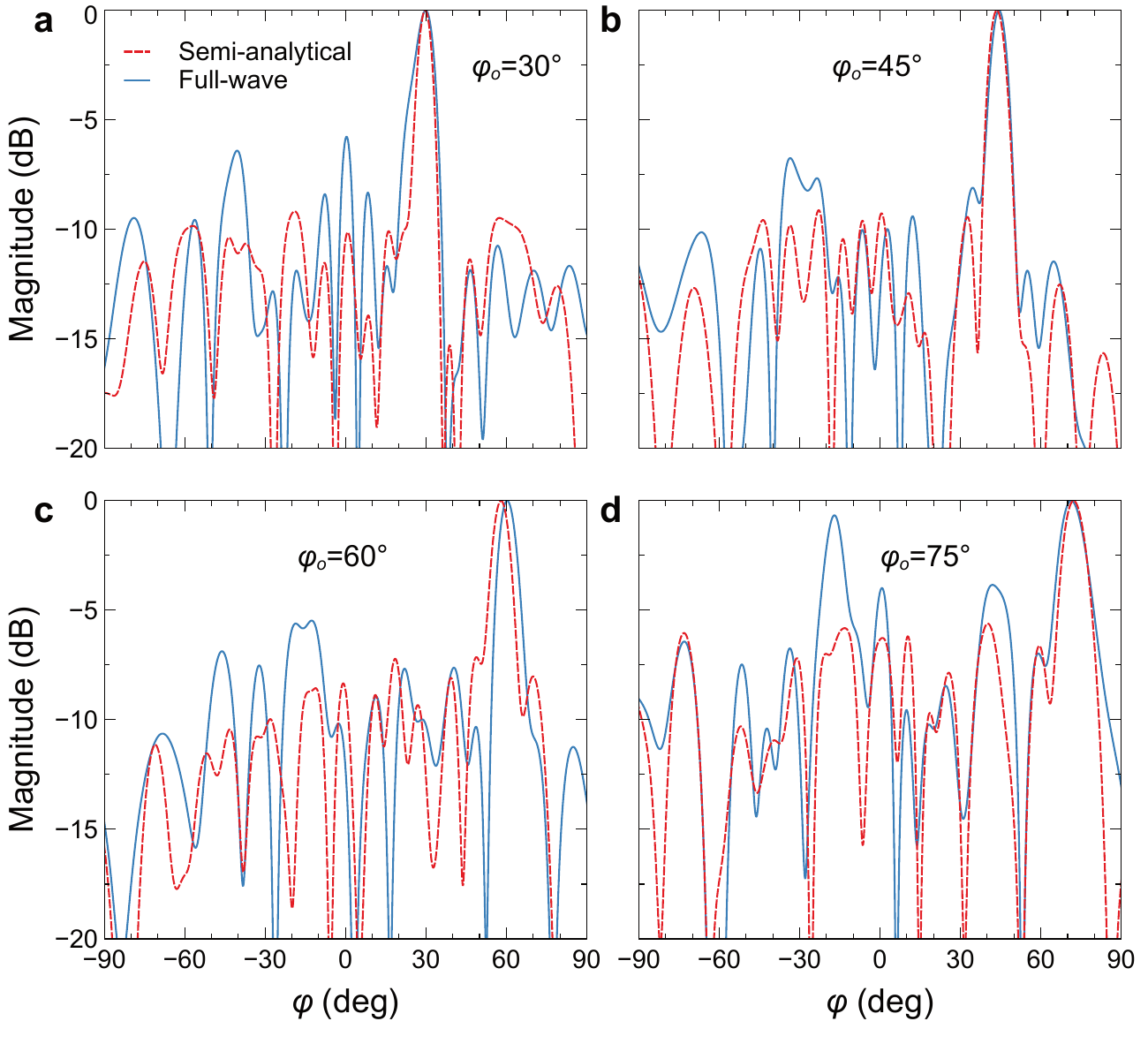}
	\caption{a,b,c,d) Comparison of scattering patterns obtained using the semi-analytical model and full-wave simulations, for  ES syntheses targeting a steering angle of $\varphi_o = 30^\circ$, $45^\circ$, $60^\circ$ and $75^\circ$, respectively. Patterns are normalized to their individual peak values.}
	\label{fig:FigureS2}
\end{figure}

Figure \ref{fig:FigureS2} presents a comparison between the scattering patterns obtained from the semi-analytical model and those computed via full-wave simulations, for ES-based syntheses targeting steering angles of $\varphi_o = 30^\circ$, $45^\circ$, $60^\circ$, and $75^\circ$, respectively.

As observed, the agreement between the semi-analytical model and full-wave simulations is generally good in the main lobe region. For the sidelobes, discrepancies remain within approximately 3 dB for steering angles of $\varphi_o = 30^\circ$, $45^\circ$, and $60^\circ$, while they increase to about 5 dB for the more extreme case of $\varphi_o = 75^\circ$. In all cases, the semi-analytical model tends to underestimate the sidelobe levels.

\begin{figure}
	\centering
	\includegraphics[width=\linewidth]{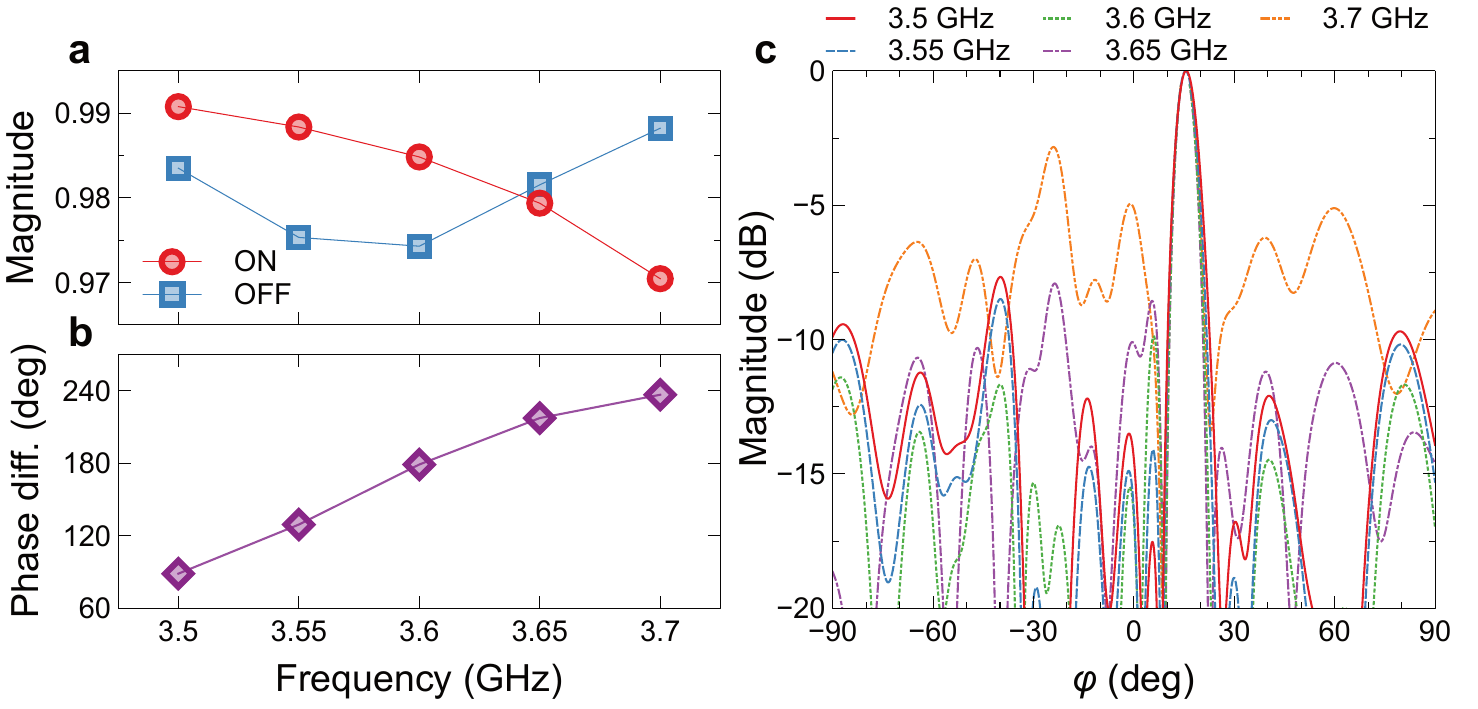}
	\caption{a, b) Magnitude and phase difference of the reflection coefficient for the one-bit reconfigurable meta-atom, evaluated at normal incidence, as a function of the operating frequency, for both operational states. c) Full-wave scattering patterns for an ES-based synthesis targeting a steering angle of $\varphi_o = 15^\circ$, shown for different frequencies around the design value. Patterns are normalized to their individual peak values.}
	\label{fig:FigureS3}
\end{figure}

Finally, Figure \ref{fig:FigureS3} illustrates the frequency sensitivity of our design. 
When the operating frequency deviates from the nominal value of 3.6 GHz, the reflection magnitude of the meta-atom in both ON and OFF states for normal incidence (Figure \ref{fig:FigureS3}a) remains relatively stable, showing only limited variations across the examined band. In contrast, the phase difference between the two states (Figure \ref{fig:FigureS3}b) exhibits a more pronounced frequency dependence and progressively departs from the ideal 180$^\circ$ value. The impact of this behavior on the overall scattering response is illustrated in Figure \ref{fig:FigureS3}c, where frequency detuning mainly results in elevated SLLs, while the direction and peak level of the main beam are largely preserved within the considered frequency range.

%

%

%

\bibliographystyle{IEEEtran}
\bibliography{C-RIS}

\end{document}

%% file: arxiv.tex
\usepackage{tikz}

\newcommand\copyrighttext{%
  \footnotesize \textcopyright \the\year{} IEEE. Personal use of this material is permitted. Permission from IEEE must be obtained for all other uses, including reprinting/republishing this material for advertising or promotional purposes, collecting new collected works for resale or redistribution to servers or lists, or reuse of any copyrighted component of this work in other works.}

\newcommand\acceptedtext{%
  \footnotesize This article has been accepted for publication in a future issue of Wiley Advanced Electronic Materials, but has not been fully edited. Content may change prior to final publication. \\
  Citation information: DOI 10.1002/aelm.202500550, Wiley Advanced Electronic Materials.
}
  
\newcommand\acceptednotice{%
\begin{tikzpicture}[remember picture,overlay]
\node[anchor=north,yshift=0pt,xshift=0pt] at (current page.north) {%
\begin{minipage}{\textwidth}
\center \acceptedtext
\end{minipage}};
\end{tikzpicture}%
}